\documentclass{emulateapj}

\slugcomment{An extreme quasar SED}

\shorttitle{The unusual SED of the quasar LBQS 0102-2713}
\shortauthors{Th. Boller, K. Linguri, T. Heftrich \& M. Weigand}

\begin{document}

\title{The unusual spectral energy distribution of LBQS 0102-2713}

\author{Th. Boller\altaffilmark{1}}
\affil{Max-Planck-Institut f\"ur extraterrestrische Physik Garching, Germany}
\email{bol@mpe.mpg.de}

\author{K. Linguri\altaffilmark{2}}
\affil{Secondary School Neu-Isenburg, Germany}

\author{T. Heftrich\altaffilmark{3}}
\affil{Johann Wolfgang Goethe-University Frankfurt am Main, Germany}

\and

\author{M. Weigand\altaffilmark{3}}
\affil{Johann Wolfgang Goethe-University Frankfurt am Main, Germany}

\begin{abstract}
\noindent
We have studied the spectral energy distribution of the
quasar~LBQS~0102-2713.
The available multiwavelength data in the observers
frame are 
one optical spectrum between 3200 and 7400 \AA\ (Morris et al. 1991), 
7 HST FOS spectra between 1700 and 2300~\AA\ (York et al. 1990), 
one {\it GALEX} NUV flux density and a $\rm K_S$ magnitude obtained from NED, 
and 3 public ROSAT PSPC pointed observations in the 0.1$-$2.4 keV energy band.
The  $\rm \alpha_{ox}$ values 
obtained from the HST FOS-, the optical spectrum, and the ROSAT observations are 
-2.3 and -2.2, respectively, comparable to BAL quasars (e.g. Gallagher et al. 
2006).  The 2500~\AA\ luminosity density is about 
a factor of 10 higher compared to the mean of the most luminous SDSS
quasars (Richards et al.  2006, their Fig. 11). 
The 2 keV $\nu L_{\nu}$ value  is lower by about a factor of 10 compared
to the radio loud quasars shown in  Fig. 10 of Richards et al. (2006). 
LBQS 0102-2713 exhibits one of the steepest soft X-ray photon indices
obtained so far. For a simple power law fit with leaving the $\rm N_H$ free
in the fit we obtain a photon index of $\rm \Gamma=6.0\pm1.3$. Fixing
the $\rm N_H$ value to the Galactic value the photon index still remains steep
with a value of about 3.5.
We argue that LBQS 0102-2713 is similar to BAL quasars with respect to
their UV brightness and  2 keV X-ray weakness. However, the absorption
by neutral matter is significantly lower compared to BAL quasars. The X-ray
weakness is most probably not due intrinsically X-ray weakness 
based on the UV line strenghts which are comparable to the line strength values
reported in quasar composites (e.g. Brotherthon et al. 2001, 
Vanden Berk et al. 2001, or Zheng et al. 1997).
If the X-ray weakness will be confirmed in future observations,
LBQS 0102-2713 might  be indicative
for a new class of quasars with an unusual combination in their
UV-, X-ray, and $\rm N_H$ properties. 
\end{abstract}
\keywords{AGN: general --- Quasars: individual LBQS 0102-2713 }


\section{Introduction}

\noindent
We report on the extreme  ultraviolet-to-X-ray spectral energy 
distribution (SED) of the quasar LBQS 0102-2713 using the spectral index 
$\rm \alpha_{ox}$. The $\rm \alpha_{ox}$ value is defined as 
$\rm \alpha_{ox}=0.384\times log\ (L_{2keV}) / (L_{2500\AA})$. The 
$\rm \alpha_{ox}$ value relates the relative efficiencies of emitted
ultraviolet disk photons to the hard X-ray photons. This value is therefore 
an important tool to provide quantitative and qualitative constrains
on models of the physical association between the UV and X-ray emission. 
The $\rm \alpha_{ox}$ value obtained from the HST FOS spectra (York et al.
1990), the optical spectrum (Morris et al. 1991), and the ROSAT data are
-2.3 or -2.2, respectively.
The $\rm \alpha_{ox}$ values are similar compared to  BAL quasars
(e.g. Gallagher et al. 2006).
The authors  have analyzed 35 BAL quasars based on Chandra
observations. Their $\rm \alpha_{ox}$ values range between -1.65 and -2.48. 
The majority of the objects have values smaller than -2.0. It is  argued 
that the X-ray weakness of BAL quasars is due to neutral intrinsic absorption 
with column densities between about $\rm (0.1-10)\times 10^{23}\ cm^{-2}$. 
As more soft X-ray photons are expected for a simple neutral absorber, 
 the absorption is assumed to be  more complex. Partial covering or 
ionized absorbers can account for this observational fact.

\noindent 
Gibson et al. (2008) have analyzed Chandra and XMM-Newton observations of 
536 SDSS quasars. They find that radio-quiet BAL quasars tend to have steeper
$\rm \alpha_{ox}$ values compared to non-BAL quasars (their Fig. 3). 
They constrain the fraction of X-ray weak non-BAL quasars and find that such 
objects are rare.  Leighly et al. (2007) report on a $\alpha_{ox}$ value 
of -2.3 in the quasar PHL 1811. Miniutti et al. (2009, submitted to MNRAS) 
found values between -1.5 and -4.3 in PHL 1092. Similar results have
been obtained by Strateva et al. (2005), Vignali et al. (2003) and Green et al.
(2008). However most of the objects with $\rm \alpha_{ox}$ values close
to -2 are upper limits. The dependence on $\rm \alpha_{ox}$  as a function
of  redshift and the rest frame ultraviolet luminosity  has 
been investigated  by Vignali et al. (2003), Strateva et al. (2005), 
and Green et al. (2008). 

\noindent
We also concentrate on the spectral energy distribution in 
the soft (0.1-2.4 keV) X-ray band based on 
ROSAT PSPC observations. We find that LBQS 0102-2713 might have an
extreme value of the photon index of $\rm \Gamma=(6.0\pm1.3)$. 
X-ray observations have shown Narrow-Line Seyfert 1 Galaxies (NLS1s) to have 
steep soft X-ray spectra as a class, first reported by Puchnarewicz et al. 
(1992). A significant correlation between the slope of the 0.1$-$2.4 keV X-ray
continua and the FWHM of the $\rm H\beta$ line was found for Seyfert 1 galaxies
(Boller et al. 1996). The distribution of the values $\rm \Gamma$ and 
FWHM $\rm H\beta$ show a continuous increase in the slope of the spectral 
continuum distribution with decreasing FWHM $\rm H\beta$ line width. 
This suggests that narrow- and broad-line Seyfert 1 galaxies form essentially 
the same class of objects and that there might be an underlying physical 
parameter which controls the distribution of objects. Smaller black hole
masses could account for narrow $\rm H\beta$ line widths which requires that 
NLS1 have to be accreting at higher fractions of their Eddington rates to 
maintain their relatively normal observed luminosities (Boller et al. 1996).
XMM-Newton observations of IRAS 13224-3809 (Boller et al. 2003) and 1H 0707-495
(Boller et al. (2003) have confirmed that a shifted accretion 
disc spectrum accounts for the steep ROSAT photon indices.
Steep soft photon indices have also been reported by other authors
(e.g. New Astronomy Reviews, 2000).
Walter \& Fink
(1993) found values ranging between 1.5 and 3.4. Fiore et al. (1998) have used 
ASCA data on the quasars NAB 0205+024 and PG 1244+026 and found photon
indices of $\rm \Gamma=(3.22\pm 0.24)\ and\ (2.95\pm 0.28)$, respectively.
George et al. (2000) report on a photon index of $\rm 4.18^{0.82}_{1.1}$ in 
PG 0003+199. Similarly to the correlation found in the soft energy range, 
a correlation between the slope of the 2$-$10 keV photon index, obtained from 
power-law fits to ASCA observations of broad- and narrow-line Seyfert 1 
galaxies, and the width of the FWHM $\rm H\beta$ line, has been discovered by 
Brandt et al. (1997). Steep 2$-$10 keV X-ray continua, with
values of the photon index between 1.9 and 2.6, are characteristic of
NLS1s. One possible explanation suggested is that NLS1s may exhibit a cooler 
accretion disc corona (Pounds et al. 1995).  A more detailed description of our 
knowledge on NLS1s is given in: The Universe in X-rays, sections 22.2 to 22.5, 
Eds. Tr\"umper \& Hasinger (2008).

\noindent
The multiwavelength  SED of radio-loud quasars has been investigated by
Richards et al. (2006), building upon the quasar SED published by
Elvis et al. (1994) for radio-quiet quasars.  Richards et al. (2006) use 
mid-infrared data based on the {\it SpitzerSpaceTelescope}, SDSS photometry 
as well as near-infrared, {\it GALEX,\ VLA} and {\it ROSAT} data. One of the 
most important results is the quasar SED shown in their Fig. 10. 
The mean radio-quiet $\rm \alpha_{ox}$ value from Elvis et al. (1994) is -1.5, 
while for radio-loud objects the mean value is -1.8 (Richards et al. 2006).

\noindent
LBQS 0102-2713 is a quasar at a redshift of 0.78. 
The object was selected from the ROSAT PSPC catalogue via the hardness 
ratio 1 (HR1)
\noindent\footnote{\sl 
The hardness ratio 1 is defined as HR1 = ((52-102)-(11-41))/((11-41)+(52-201))
Zimmermann et al. 1994. The numbers refer to the ROSAT PSPC channels.
}
criteria in the range from -0.5 to -0.7 derived from a reference 
sample of known steep spectrum AGNs. LBQS 0102-2713 showed the steepest photon 
index from the selected sample. Here we concentrate on 3 public ROSAT PSPC 
observations, an optical spectrum between the 3200 to 7400 \AA\ observed in 
1988 (Morris et al. 1991), 7 HST FOS spectra between 1700 and 2300 \AA, 
and one {\it GALEX} NUV flux density available at NED. 
High energy data above 2.4 keV are not available presently. In Section 2 we 
describe the X-ray observations and data analysis. The results from the X-ray 
fitting analysis is given in Section 3. Section 4 contains the results from 
the $\rm \alpha_{ox}$ analysis. The comparison with mean SED's is given in 
Section 5. Models for the X-ray weakness are presented in Section 6.
Section 7
contains the summary and open problems. In Section 8 we list security checks
on the identification of LBQS 0102-2713 as a X-ray source.
Throughout the paper we adopt $\rm H_0=70\ km\ s^{-1}$ and a $\rm
\Lambda$-cosmology of $\rm \Omega_M=0.3\ and\ \Omega_{\Lambda}=0.7$.

\section{X-ray observations and data analysis}

\noindent
The object was observed three times with the ROSAT PSPC detector in 1992 
(c.f. Table 1 for the observation log file).
The source is located off-axis in the PSPC detector with a separation 
from the pointing direction of about 0.5 degrees.
The data were retrieved from the MPE archive. 
The PSPC data in the MPE archive 
were processed at MPE and copied to HEASARC. The ROSAT data in both 
archives are identical.   

\noindent
The ROSAT data were converted into FITS files using the {\it xselect}
command-line interface version 2.4. The observations were converted 
into FITS files and a merged FITS file was created from the two longer
observations (ROR numbers 700121p-0 and 700121p-2). The short observation
(ROR number 700121p-1) is highly background dominated and the source is not
significantly detected. Therefore the results from the spectral fitting 
are only presented for the two long observations and the corresponding
merged data set. The {\it xselect} command-line interfaces handles RDF
FITS format files. In the RDF case the events are for instance in the file 
rp700121a00\_bas.fits file (the basic file) for ROR number 700121p-0.
Using the {\it read events}
command the two single observations and the merged observation the were read
into {\it xselect}. ROSAT images were created via {\it extract image} and
{\it saoimage}. For all three observations used for subsequent spectral
fitting the source region was extracted at $\rm RA=01^h 04^m 40.9^s$ and 
$\rm DEC=-26^{o} 57^m 07^s$.  
The background region was extracted at
$\rm RA=01^h 04^m 09^s$ and 
$\rm DEC=-26^{o} 45^m 04^s$. The source and background
extraction radii are 320 arcsec.
Finally using {\it extract spectrum} and {\it save spectrum} the source and
background fits files were created. The corresponding ROSAT  response matrix 
is {\it pspcb\_92mar11.rsp}.

\noindent
We have analyzed the ROSAT PSPC observations of LBQS 0102-2713
independently from the {\it xselect} command-line interface with the 
{\it EXSAS} software routines (Zimmermann et al. 1994) developed for
analyzing ROSAT data based on the {\it MIDAS} software package for
the two long observations
700121p-0 and 700121p-2. 
A merged data file was also created with the
{\it EXSAS} command {\it intape/disk}. 
Energy channels were selected from 8 to 240 using the 
{\it EXSAS prepare-spectrum} programme.

\section{Spectral fitting}
\noindent
In the ROSAT All-Sky Survey (RASS) we get 26 source counts with a mean exposure 
of 333 seconds. The extraction radius is 500 arcsec. For the background we 
get 16 counts with a mean exposure of 435 seconds and an extraction radius of 
500 arcsec. This results in 10 net counts and a mean count rate of 
$\rm 0.02\ counts\ s^{-1}$. With these numbers no reliable fits could be 
obtained from the RASS data. For the merged data sets of the PSPC pointings 
we get 1351 source counts and 741 background counts for a total exposure of 
12881 seconds. This results in 610 net counts and a mean count rate of 
$\rm 0.05\ counts\ s^{-1}$. With these numbers and the limited spectral energy 
resolution of ROSAT only a limited range of models can be fitted. 
We present the results only for a simple power law fit with cold absorption 
and a disk black body model. Longer X-ray observation with the present 
generation of X-ray satellites are required to present more advanced model 
fits, 
e.g. a smeared absorption model according to Schurch \& Done (2006), or
 a disk reflection model following (Crummy et al. 2006).

\subsection{xselect command-line interface and EXSAS spectral power law fitting}
\noindent
The source spectra obtained from the ROR numbers 700121p-0 and 700121p-2 using 
the {\it xselect} command-line interface were grouped with the {\it grppha} 
command with {\it group min 10}. The merged data set was grouped 
with {\it group min 30}. The spectral fitting was performed using {\it XSPEC} 
version 12.5.0. The model components were (mo~=~phabs~(zpo)).
While the light curves do not show significant variations (Fig. 1) the 
resulting photon indices are remarkably steep (Fig. 2). LBQS 0102-2713 might 
be the quasar with the steepest soft X-ray photon index, detected in two
individual and the merged observations,  reported so far. For the merged data 
set we get $\rm \Gamma=(6.0\pm 1.3)$ and $\rm \Gamma=(5.8\pm 1.3)$ for the 
{\it xselect} and {\it EXSAS} software packages, respectively. 
The spectral fitting results from the {\it xselect} command line interface and
 the 
{\it EXSAS} software routines are listed in Table 2.
The spectral parameters obtained from both software systems are consistent 
within the errors. As a security check we have used the NASA GSFC simulation 
software {\it webpimms}. The simulated spectra are consistent with the ROSAT 
power law fits. This is an independent test, whether the calibration and 
response files available at the MPE site give the same results as the 
corresponding files produced by the NASA GSFC calibration team.

\subsection{$\rm Gamma-N_H$ contour plots}
\noindent
The Galactic foreground absorption in the direction to LBQS 0102-2713
is very low with $\rm N_H=1.2\times 10^{20}\ cm^{-2}$. It is known
that $\rm \Gamma$ and $\rm N_H$ are correlated in ROSAT PSPC fits, such
that a larger $\rm N_H$ requires a steeper photon index to give a good fit.
In Fig 3 we show the $\rm \Gamma-N_H$ contour plots for the merged data set
and for the ROR number 700121p-2. For the ROR number 700121p-0 no reliable 
contour plot could be obtained. At a 99 per cent confidence level the
the photon index is about 3.5 at the Galactic $\rm N_H$ value. 
The photon index might indeed be flatter as 
$\rm \Gamma=(6.0\pm1.3)$ and the object may be less special in X-rays than it 
appears from the simple power law fits. However, the photon index obtained for
a power law fit with the Galactic  $\rm N_H$ value fixed remains still steep 
for 
quasar soft X-ray spectral energy distributions (c.f. Section 1). Longer
observations with the present generation of X-ray telescopes are required to
confirm whether the photon index of LBQS 0102-2713 is indeed extremely steep.

\subsection{Ionized absorber plus power law spectral modelling}
\noindent
We note that a simulation  of a power law  
plus  an ionized absorber could also result into a flatter photon index.  
A simulated  50 ks XMM-Newton spectrum obtained from a power law model
with cold absorption plus an ionized absorber based on the 
{\it grid25BIG\_mt.fits} model (Done, private communication)
gives a photon index of $\rm \Gamma=(4.4\pm0.5)$ for a column density of
the ionized absorber ranging between  $\rm (1.1-1.5)\times 10^{22}\ cm^{-2}$ 
and an ionization parameter range between $\rm \xi = (20-30)$.
The parameter ranges explored range between $\rm N_H=10^{(21-24)}\ cm^{-2}$ and
$\rm \xi =$ 10 to 1000. The flatter photon index occurs in a very narrow
parameter range of the ionized absorber column density and the ionization
value. 
When fitting the simulated data with the ROSAT PSPC response and the ROSAT
exposure time, the photon index is $\rm (6.7\pm1.8)$,
consistent with the ROSAT spectral fitting results for a power law model
with the $\rm N_H$ value leaving free in the fit.

\subsection{Black body spectral fitting results}
\noindent
A black body fit to the merged data set does also result into an acceptable 
fit. The derived spectral parameters in the rest frame are:
$\rm N_H=(2.0\pm2.3)\times 10^{20}\ cm^{-2}$,
$\rm kT=(0.13\pm0.013)\ keV$, and
$\rm n=(2.6\pm2.8)\times 10^{-5}\ photons\ cm^{-2}\ s^{-1}\ keV^{-1} $ at 
1 keV. Due to the limited spectral resolution of ROSAT 
we cannot disentangle between the a power law and a black body model.
We compare the derived black body temperature with other published black body
temperatures and find that  they are in good agreement.
Crummy et al. (2006) found values for the black body
temperature for an AGN sample  observed with XMM-Newton between 0.009 and 
0.17 keV (their Table 2). Tanaka, Boller and Gallo (2005) 
have analyzed NLS1 spectra observed with XMM-Newton and found for a sample of 
17 objects black body temperatures between 0.09 and 0.15 keV.
Fiore et al. (1998) obtain black body temperatures of 0.16 keV
for PG 1244+026 and NAB 0205+024, respectively  based on ASCA observations. 
Puchnarewicz et al. (2001) obtain a value of 0.13 keV for the 
NLS1 galaxy RE J1034+396.

\section{Deriving the $\alpha_{ox}$ value for LBQS 0102-2713}

\subsection{Deriving the 2500 \AA\ flux density from the optical spectrum}
\noindent
Morris et al. (1991) have published an optical spectrum of LBQS
0102-2713 in the wavelength  range between 3200 and 7400 \AA\ in the observers 
frame (c.f. Fig. 4). The  Mg II line at 2800 \AA\ (rest frame) is clearly 
visible. Applying a Gaussian fit to this line we obtain a FWHM value of about 
2200 km $\rm s^{-1}$ in the rest frame. This value is  very close to the 
artificial border line between NLS1 galaxies and broad line Seyfert galaxies 
of 2000~ km~$\rm s^{-1}$  following the  definition of Osterbrock and Pogge 
(1985). 
In addition Morris et al. (1991) first noted the strong UV Fe II multiplet 
emission between about 2200 and 2500 \AA\ in the
rest frame. All this is typical for NLS1 
galaxies and LBQS 0102-2713 can be considered as a X-ray bright NLS1
galaxy. This goes in line with the steep X-ray spectrum.
In addition a strong C III] line is found  at 1909 \AA\ rest frame and a 
strong emission feature at about 2075 \AA\ which is usually interpreted as a 
bunch of Fe III multiplets. The later feature  is catalogued in the Vanden 
Berg quasar composite paper (2001)  and is certainly present in 
higher signal-to-noise composites.

\noindent
The observed 4450 \AA\ flux density from the optical spectrum (c.f. Fig. 4) 
obtained in 1988 is about $\rm f_{4450 \AA }^{obs.} = 3.0\times
10^{-16}\ erg\ cm^{-2}\ s^{-1}\ \AA ^{-1}$.
Details of the spectroscopic measurements can be found in section 2.3 of
Morris et al. (1991). To convert the units from \AA\ to Hz we followed
the relation from Hogg (2000) described in his Chapter 7. 
As the differential flux per unit log frequency or log wavelength is
$\rm \nu f_{\nu}=\lambda f_{\lambda}$ (Hogg uses S instead of f) one gets
$\rm f_{\nu}= (\lambda^2 / c)\times f_{\lambda}$ with $\rm f_{\lambda}=
f_{4450 \AA}^{obs.}$ and $\rm \lambda= 4450\ \AA$. The resulting observed flux 
density per Hz is $\rm f_{4500 \AA}^{obs.}= 2.0\times 10^{-27}\ erg\ cm^{-2}\ 
s^{-1}\ Hz^{-1}$.  To correct the monochromatic flux measurements for Galactic 
extinction we used the Nandy et al. (1975) extinction law with 
$\rm R=A_V/E(B-V)$. Their Table 3 gives for $\rm (1/ \lambda (\mu\ m^{-1}))$ 
at 4450 \AA\ a R value of 3.8.
The E(B-V) value from the GALEX data obtained from NED is 0.02.
According to Geminale \& Popowski (2005) the relation between R and 
$\rm \epsilon$ is $\rm \epsilon(\lambda\ - V) = R((A_{\lambda} / A_V) - 1)$. 
Following their extinction curves shown in Fig. 4 one obtains an 
$\rm \epsilon$ value of about 0.5. The $\rm A_{\lambda}$ value at 
4450 \AA\ is then 0.076 and  the $\rm A_{V}$ value using the relations given 
above is 0.066. The $\rm A_{\lambda} / A_{V}$ value is  1.15.
The extinction corrected flux and the uncorrected flux at 4450 \AA\ 
are related by $\rm f_{4450 \AA}^{Ext.} = (10^{+4}\times A_{\lambda} / A_V)\times
f_{4450 \AA}^{obs.} = 2.6\times f_{4450 \AA}^{obs.}= 5.3\times
10^{-27}\ erg\ cm^{-2}\ s^{-1}\ Hz^{-1}$. 
The luminosity distance $\rm D_L= 1.5\times 10^{28}\ cm$.
The resulting rest frame UV luminosity density  
is $\rm L_{2500 \AA }= 1.6\times 10^{31}\ erg\ s^{-1}\ 
Hz^{-1}$. The relation between $\rm l_{uv}$ and $\rm L_{2500 \AA}$ is
$\rm log\ L_{2500 \AA} = l_{uv} = 31.21$.

\subsection{Deriving the 2500 \AA\ flux density from the HST FOS spectra}
\noindent
We also analyzed the 7 HST FOS spectra named as Y309010nT with n ranging from 2 
to 8.  The HST FOS spectra were obtained from the HST FOS proposal ID
6007:  Comparison of the large scale structure in QSO absorbers and galaxies 
in the  Galactic Poles, D. York, University of Chicago. The observed
wavelength ranges between 1700 and 2300 \AA. All 7 exposures obtained in 1995
agree in their flux density values and the HST FOS absolute and relative 
calibration is very good (M. Rosa, ESO Garching, private communication). 
In Fig. 5 we show the spectrum for the science exposure number Y3090102T.
The $\rm Ly_{\alpha}$ line is clearly visible and marked in the spectrum.
The line is blended with a low ionization line of N V at 1240 and 1242
\AA\ in the rest frame. In addition there is emission from 
the $\rm Ly_{\beta}$ line blended with the O VI line at 1032 and 1037 \AA\ at 
their rest frame wavelengths.

\noindent
The observed 2300 \AA\ flux density is about
$\rm f_{2300 \AA}^{obs.} = 1.5 \times 10^{-15}\ erg\ cm^{-2}\ s^{-1}\ \AA ^{-1}$.
The flux density at 2500 \AA\ rest frame is 
$\rm f_{2500 \AA} = f_{1292 \AA}\times (1292 \AA\ / 2500 \AA\ )^{-0.5} =
4.2\times 10^{-27}\ erg\ cm^{-2}\ s^{-1}\ Hz^{-1}$.
To extrapolate the corresponding rest frame wavelength of 1292 \AA\ to
2500 \AA\ we have used the $\rm \alpha_o$ value of -0.5 of Richstone \&
Schmidt (2000).
The extinction corrected flux at 2500 \AA\ is 
$\rm 1.4 \times 10^{-26}\ erg\ cm^{-2}\ s^{-1}\ Hz^{-1}$, about a factor of
3.3 larger compared to the non-extinction corrected flux density. The
slight discrepancy compared to the optical spectrum obtained by Morris et
al. (1991) might be due to flux variability between the 1988 and 1995 observations
or due to the extrapolation from 1292 to 2500 \AA\ using a mean 
$\rm \alpha_o$ value of -0.5. 
The correction for Galactic extinction has been performed as described 
in the previous subsection.
The R value at 2500 \AA\ is 7.2 and the $\rm \epsilon$ value is 7.2. 
The resulting $\rm A_{\lambda} / A_{V}$ value is 1.4. This results 
into a Galactic extinction correction with a factor 3.3 and the 
extinction corrected flux is 
$\rm f_{2500 \AA}^{Ext.} = 1.4\times 10^{-26}\ erg\ cm^{-2}\ s^{-1}\ Hz^{-1}$. 
The resulting rest frame luminosity density 
is $\rm L_{2500 \AA } = 4.0\times 10^{31}\ erg\ s^{-1}\ Hz^{-1}$.

\subsection{Deriving the 2 keV flux density from the merged ROSAT PSPC data set}
\noindent
The 2 keV rest frame flux density is determined following Hogg (2000) and
Weedman's Quasar Astronomy (Section 3.5, page 61, 1986).
The photon index $\rm \Gamma$ is related to the energy index via 
$\rm \alpha_x = -\Gamma + 1$.
For a photon index not equal to 2 the rest frame 2 keV flux density is
given by
$\rm f_{2keV} = f(0.5-2.0)\times ((1+\alpha_x) / (1+z)^{\alpha_x}))\times
((\nu_{2keV}^{\alpha_x} / (\nu_{2keV}^{\alpha_x + 1} - \nu_{0.5keV}^{\alpha_x +
1}))$.  
The  unabsorbed ROSAT flux in the 0.5-2.0 keV energy band is
$\rm 4.3\times 10^{-14}\ erg\ cm^{-2}\ s^{-1}$. As the photon index derived
in the previous section was 6.0, we get for $\rm \alpha_x$ a value of -5.
The corresponding 2 keV and 0.5 keV frequencies are
$\rm 4.8\times 10^{17}\ Hz\ and\ 1.2\times 10^{17}\ Hz$. This results
into a flux density at 2 keV of
$\rm f_{\nu\ 2keV} = 4.3\times 10^{-32}\ erg\ cm^{-2}\ s^{-1}\ Hz^{-1}$. 
The 2 keV luminosity density is then 
$\rm L_{2keV} = 4.0 \times 10^{25}\ erg\ s^{-1}\ Hz^{-1}$.
We note that the extremely steep photon index has to be confirmed in future 
observations  with the present generation of X-ray telescopes given the results 
from the contour plots shown in Fig. 3. If the X-ray spectrum is flatter than 
obtained  from the power law fits, this would result in a less extreme
 $\alpha_{ox}$ value.

\noindent
For a definition of 
$\rm \alpha_{ox} = 0.384\times log (L_{2keV} / L_{2500\AA })$ one derives
$\rm \alpha_{ox}$ values of -2.3 and -2.2.

\section{Comparison with quasar SED's}
\subsection{Comparison with mean SDSS quasar SED's}
\noindent
Richards et al. (2006) show the mean quasar SED's for optically luminous SDSS-, 
all SDSS-, and optically dim quasars (c.f. Fig. 11 of their paper). 
In Fig. 6 we show an adopted version of the plot of Richards et
al. (2006). The $\rm \nu L_{\nu}$ 
value at 2500 \AA\ rest frame wavelength is $\rm 4\times10^{47}\ erg\ 
s^{-1}$. This is about a factor of 10 more luminous compared to the mean of the 
optically most luminous SDSS quasars. However there is a large dispersion in 
the optical $\rm M_B$ magnitudes for quasars. 
Brotherton et al. (2001) show the distribution of the $\rm M_B$ magnitudes for more than 
11000 quasars (their Fig. 5). Their absolute magnitudes range between about $\rm -23^{mag}$ to
$\rm -32^{mag}$. LBQS 0102-2713 has a photometric IIIa-J magnitude of 17.52 mag as taken
from NED at a wavelength of 4612 \AA\ corresponding to 0.46 $\rm \mu$m. 
This close to the mean B wavelength of 0.40 $\rm \mu$m. The corresponding 
$\rm M_B$ value  of LBQS 0102-2713 is -25.9. This might indicate that the 
scatter in the 2500 \AA\  luminosities is also large and that 
LBQS 0102-2713 is less extreme in the UV than one would think at first glance 
when comparing the UV value to the mean of  the optically most luminous quasars. 
The corresponding 2.0 keV rest frame value of $\rm \nu L_{\nu}$ value is 
$\rm 1.4\times10^{44}\ erg\ s^{-1}$ which is comparable to the optically 
dim SDSS quasars. The dispersion in the X-rays appears to be much less compared
to the UV luminosities if one compares the X-ray luminosity of LBQS 0102-2713 
to the X-ray luminosities as shown in the sample papers of Strateva et al. 
(2005), Vignali et al. (2003) or Gibson et al. (2008).

\subsection{Comparison with the quasar SED for radio-loud quasars}
\noindent 
In Fig. 7 we have plotted the available multiwavelength data in units
of $\rm \nu f_{\nu}$ versus the wavelength and compare these data points
to the mean quasar SED for radio-loud objects from Richards et al. (2006, 
their Fig. 10).
All data points are extinction corrected. For the ROSAT data we show only the
lowest and highest energy data point. While at 2 keV the object is X-ray weak 
compared to the mean quasar SED, at the softest energies LBQS 0102-2713 is a
very powerful emitter and the $\rm \nu f_{\nu}$ value even exceeds the 2500 \AA\
UV value. The HST FOS data points are about a factor of 2 to 5 larger compared
to the mean quasar SED. Here we note again that the scatter in the individual 
UV luminosities compared to the mean might be larger as pointed out in the 
previous subsection and that therefore the object might be less extreme in it's UV 
luminosity. The optical data points and the $\rm K_S$
magnitude follow the mean quasar SED.
The main result from this plot is the  discrepancy between the X-ray
and HST FOS data points as these bands are discontinuous. 
The standard picture assumes that the rest frame UV photons are emitted from
the accretion disc and that the hard X-ray photons are arising from the
accretion disc corona. From Fig. 7 it is obvious that there is a discrepancy 
between the UV and the X-ray photons.
The optical data from Morris et al. (1991) are  also offset with respect
to the HST FOS data points. This is most probably due to a significant 
contribution from the host galaxy and the rest frame optical photons are not 
expected to be related to the accretion disc.
We have cross-checked the discontinuity between the UV and X-ray data points.
The GALEX NUV wavelength is 2315.7 \AA\ (c.f. Morrissey et al. 2007, their Table 1). 
The NUV GALEX calibrated flux from NED is 
282.2 $\rm \mu$Jy. This converts into  flux densities of 
$\rm 2.8\times 10^{-27}\ erg\ cm^{-2}\ s^{-1}\ Hz^{-1}$
or $\rm 1.6\times 10^{-15}\ erg\ cm^{-2}\ s^{-1}\ \AA^{-1}$. 
The GALEX NUV flux density value is consistent with the flux 
density shown in the HST FOS spectrum is
Fig. 5.

\section{Models for the X-ray weakness}

\subsection{The intrinsically weakness model}

\noindent
Leighly et al. (2007) favor an intrinsically 
X-ray weakness model for the quasar PHL 1811. The 2 keV X-ray emission is 
intrinsically weak rather than absorbed. In this case one expects weak 
low ionization of semiforbidden UV lines. For the 
blend of the $\rm Ly_{\alpha}$ and N~V lines an EW value of 
15 eV is obtained. No EW values are given for
the blend of $\rm Ly_{\beta}$ and O VI.
The authors argue that this points to an intrinsically weak X-ray model. 
The argumention for an
intrinsically  X-ray weakness model is based on individual lines such as 
Na I D, Ca II H and
K and C IV.
The rest frame EW values for the blend of 
$\rm Ly_{\beta}$ and the O VI lines and the 
$\rm Ly_{\alpha}$ and the N V lines shown in Fig. 5 are
about 12 and 50 \AA, respectively. 
In the following we compare these values with quasar composite spectra.
Brotherton et al. (2001) give for the $\rm Ly_{\beta}$ plus O VI lines an 
EW value of 11 \AA\ in the rest frame. For the blend of 
$\rm Ly_{\alpha}$ and N V the EW value is 87 \AA .
The corresponding values reported by Vanden Berk et al. (2001) 
are 9  and 94 \AA\, respectively.    
Zheng et al. (1997) found values of 16 and  102 \AA\, 
respectively.
The EW values obtained for LBQS 0102-2713 are typical to the
mean quasar composite values and the source appears not to be
intrinsically X-ray weak.

\subsection{Other models}

\noindent
There are other models in the literature which are speculative or do have a low
probability to explain the X-ray weakness.

\noindent
Gibson et al. (2008) present in their Fig. 2 a decreasing trend of 
$\rm \alpha_{ox}$ with increasing 2500 \AA\ UV luminosity density. 
This is a strong observational constraint as an 
increased UV luminosity density results into steeper $\rm \alpha_{ox}$ values. 
In the case of PHL 1811 and LBQS 0102-2713 we   indeed see lower 
2 keV flux densities compared to other quasars. 
If one compares the UV and X-ray luminosity densities of LBQS 0102-2713 
which are $\rm log\ l_{UV}= 31.20$ for the Morris et al. (1991) spectrum and
$\rm log\ l_{UV} = 31.60$ obtained from the HST FOS spectra 
and an X-ray luminosity
density of $\rm log\ l_X=25.60$ obtained from ROSAT, with Fig. 8 of Vignali et al. (2003)
the X-ray weakness becomes immediately apparent.
The reason for the
so called global X-ray Baldwin effect is presently not known. It
is speculated that a patchy or disrupted accretion disc corona is related to 
the UV luminosity density (Gibson et al. 2008).
However there is no direct observational proof for this model.

\noindent
One could speculate that
UV and X-ray variability of the source might account for the 
low $\rm \alpha_{ox}$ value. 
Assuming that LBQS 0102-2713 was in 1992 in a low X-ray
flux state, then a factor of about 10 at 2 keV is required to obtain the 
canonical  $\rm \alpha_{ox}$ value of -1.8 for radio-loud quasar as shown in 
Fig. 11 of Richards et al. (2006). A factor of 10 in three years with 
respect to the optical observations  appears unlikely as only a few AGN are 
known to exhibit flux variability of a factor  of 10 or more. 
In addition, variability seems unlikely given 
Hook et al. (1994)
 who find
only weak optical variability with $\rm \sigma$=0.17.

\section{Summary and open problems}

\subsection{Comparison to BAL quasars and a possible unique combination of 
UV brightness, X-ray weakness and low $\rm {\bf N_H}$ values}
\noindent
LBQS 0102-2713 appears to be very similar to BAL quasars which show strong 
UV- and weak 2 keV X-ray emission (c.f. Gallagher et al. 2006).
Their $\rm \alpha_{ox}$ values range between -1.65 and -2.48. The majority of
the objects have values smaller than -2.0. They authors  argue that the X-ray
weakness of BAL quasars is due to neutral intrinsic absorption with column 
densities
between about $\rm (0.1-10)\times 10^{23}\ cm^{-2}$. As more soft X-ray photons
are expected for a simple neutral absorber, the authors argue that the 
absorption
is more complex. Partial covering or ionized absorbers can account for this
observational fact. 
However, LBQS 0102-2713 do not show such high values for the neutral absorber 
compared to BAL quasars. The foreground absorption is about 
$\rm 6\times 10^{20}\ cm^{-2}$. 
This is at least a factor of about 20 lower compared to BAL quasars. 
LBQS 0102-2713 is an 
object which exhibits an unusual combination of UV brightness, 
X-ray weakness and no significant absorption by neutral matter along
the line of sight. In addition, 
there are no significant indications that the
object is intrinsically X-ray weak, in contrast to the argumentation for
PHL 1811 by Leighly et al. (2007).
This parameter combination is new and needs to be explained. 
With the present available
X-ray data we are limited in providing self consistent models but would like 
to 
present these new results to the community.

\subsection{LBQS 0102-2713 as a supersoft quasar}
\noindent
The object might exhibit the steepest photon 
index reported from a quasar so far. For a simple power law fit with neutral 
absorption  left free in the fit we get photon indices of 
$\rm \Gamma = (5.8\pm1.3)\ or\ (6.0\pm1.3)$ by using the 
{\it xselect} command line 
interface and the {\it EXSAS} software package, respectively. 
However, if the foreground
$\rm N_H$ value is set to the Galactic value of $\rm 2\times10^{20}\ cm^{-2}$
(Dickey \& Lockman, 1990) 
the photon index is about 3.5 (c.f. the $\rm \Gamma\ - N_H$ contour plots shown in 
Fig. 3). The photon index still remains  steep compared to previous 
studies listed in Section 1.  
In addition, 
Puchnarewicz et al. (1995) applied  a broken power law fit
to RE J2248-511 with a photon index of $\rm 4.13^{+5.85}_{-0.60}$ up to the break
energy of 0.26 keV.
The Galactic $\rm N_H$ value was fixed to $\rm 1.4\times10^{20}\ cm^{-2}$.
 For the object RE J1034+396 Puchnarewicz et al. (1995)
obtain for their broken power law fit a value of 
$\rm \Gamma = 
4.45^{+0.25}_{-0.32}$ up to a break energy of 0.41 keV. The {$\rm N_H$} value
was fixed to the Galactic value of $\rm  1.5\times10^{20}\ cm^{-2}$.
We note that the low $\rm N_H$ values, although fixed to their Galactic values,
result in similar steep photon indices as obtained for LBQS 0102-2713
with the $\rm N_H$ value fixed in the fit.

\subsection{Open problems}
\noindent
There are open problems which cannot be solved given the present set of 
multiwavelength
data.  First, we found that the UV emission is discontinuous to the X-ray 
emission, e.g.
the UV photons which are expected to arise from the accretion disc 
appear not to be  
correlated with  the X-ray photons. Second, if the steep soft photon index 
of  $\rm (6.0\pm1.3)$ will be  confirmed by new X-ray observations 
this steepness needs to be 
explained.  
Based on  the contour plots shown in Fig. 3, the object might not be as 
extreme as one would
expect from the simple power law fits.
Third, the UV luminosity density is about a factor of 10 more luminous 
compared to the mean of the most luminous SDSS quasars. However, as pointed out
there is
a large spread in the $\rm M_B$ magnitudes for quasars and 
the UV brightness might also show a large scatter compared to the mean value
at a certain frequency.
Finally, if the 2 keV X-ray weakness is confirmed by other observations, the
combination of UV brightness, X-ray weakness and the absence of significant 
absorption by neutral matter results into an unusual combination observational
parameters which cannot be explained with the available set of multiwavelength
data.
Finally, what is the spectral complexity in the soft and hard X-ray band 
unresolvable
for ROSAT? We know from observations with the present generation of X-ray 
telescopes that new spectral components can often be added to the fit and that 
the real spectrum might 
be more
complex. We expect to see this in the soft band with higher signal-to-noise
observations.  In addition,
observations above 2.4 keV are required to better constrain physical
models for the unusual observational parameters detected in LBQS 0102-2713.

\section{Security check}
\noindent
The probability that the X-rays arise from nearby sources is low.
The total 1 $\rm \sigma$ position error of the ROSAT detection is 7 arcsec including a 6
arcsec systematic error (see Voges et al. 1999 and the link to the source catalogue
entries).
The nearest SIMBAD source has a distance
to the position of LBQS 0102-2713 of 84 arcsec resulting into a discrepancy 
of 12 $\rm \sigma$ in the
position offset.  The same holds for the NED detections.
The next source is at a distance of 0.8 arcmin resulting into a position 
error offset of about 7 $\rm \sigma$.
In Fig. 8 we show the
optical image with the SIMBAD and NED detections overlayed. In Tables 3 and 4 
we list 
objects within a distance of 500 arcsec to the position of LBQS 0102-2713.

\acknowledgments
\noindent
The paper is based on observations obtained with the ROSAT PSPC satellite.
We thank the anonymous referee for his/her extremely helpful comments to
improve the scientific content and the structure of the paper.
TB is grateful for intensive discussion with Ari Laor and Narum Arav on
the UV- X-ray relations in quasars and especially on the object properties presented
in this paper. TB acknowledges many  comments from Niel Brandt on the UV- 
and X-ray properties reported in this paper and on his intensive suggestions with
respect to the comparison to BAL quasars. 
TB thanks Chris Done for providing her ionized absorber model and 
many suggestions to improve the paper.
TB is also grateful to Gordon Richards for providing the super mongo script
for the SED of radio-loud and radio-quiet quasars. 
TB would like to thank Michael Rosa for his information regarding the
HST FOS flux calibration for LBQS 0102-2713 and Don Neill for his help
to achieve precise information regarding the GALEX data.
KL is grateful to the Secondary School Neu-Isenburg for their 
support to work together with TB at MPE Garching in analyzing the
multiwavelength
data presented in this paper.
TH greatly acknowledges the collaboration with TB and WG in writing up
this paper. 
The authors acknowledge Iskra Strateva, Frank Haberl, Marcella Brusa and Konrad Dennerl 
for many helpful suggestions and information on the data presented in this 
paper.

\noindent
{\it Facilities:} \facility{ROSAT, HST FOS, GALEX}.

\clearpage



\noindent
\clearpage
\begin{table}
\begin{center}
\caption{Observation log data of the ROSAT PSPC observations of LBQS 0102-2713}
\begin{tabular}{lll}
\tableline\tableline                                                                           
       ROR number                &  Date           &   Observation duration [sec] \\
       700121p-0                 &January 5, 1992  &   6157                       \\
       700121p-1                 &June 5, 1992     &   2191                       \\
       700121p-2                 &December 8, 1992 &   6724                       \\
\tableline
\end{tabular}
\end{center}
\end{table}
\clearpage

\noindent
\clearpage
\begin{table}
\begin{center}
\caption{Spectral fitting results for a power law model with foreground 
absorption of $\rm LBQS\ 0102-2713^a$}
\begin{tabular}{lllllllll}
\tableline\tableline                                                                 
               &                &  {\it xselect}  &                   &              &   & {\it EXSAS}   &               &                \\
	       & $\rm \Gamma$   & $\rm N_H$       & norm              &              &   &$\rm \Gamma$   & $N_H$         & norm           \\
ROR number     &                &                 &                   &              &   &               &               &                \\
700121p-0      & $\rm 6.6\pm2.5$&$\rm 6.2\pm0.5$  &$\rm 2.5\pm2.9$    &              &   &$\rm 5.5\pm2.5$&$\rm 7.4\pm1.5$&$\rm 4.2\pm5.0$ \\
700121p-2      & $\rm 5.5\pm2.5$&$\rm 3.7\pm2.4$  &$\rm 1.3\pm1.8$    &              &   &$\rm 6.5\pm2.1$&$\rm 6.0\pm4.7$&$\rm 4.5\pm4.0$ \\
merge data set & $\rm 6.0\pm1.3$&$\rm 4.8\pm1.5$  &$\rm 1.3\pm1.6$    &              &   &$\rm 5.8\pm1.3$&$\rm 6.5\pm4.4$&$\rm 1.5\pm1.0$ \\
\tableline
\end{tabular}
\tablenotetext{a}{The $N_H$ value is in units of $\rm 10^{20}cm^{-2}$ and
the normalization is given in $\rm 10^{-4}\ photons\ cm^{-2}\ s^{-1}\ keV^{-1}$ at 1 keV.}
\end{center}
\end{table}
\clearpage

\begin{table}
\begin{center}
\caption{SIMBAD detection within a  500 arcsec radius around LBQS 0102-2713\label{tbl-2}}
\begin{tabular}{llllc}
\tableline\tableline
Identifier            & dist(arcsec) & type  & ICRS (2000) coord.       & Sp type \\ 
\tableline
HB89 0102-272       & 0.00         & QSO   & 01 04 40.94  -26 57 07.5 &     -   \\
1RXS J010441.1-265712 &  5.44        &   X   & 01 04 41.10  -26 57 12.5 &     -   \\             
PHL 7126              &  84.37       &   blu & 01 04.60.00  -26 58 00.0 &     -   \\         
675                   & 148.54       &   WD* & 01 04 47.00  -26 59 12.0 &     -   \\
GEN  +6.20077018      & 184.71       &   *   & 01 04.60.00  -27 00 00.0 &     -   \\
GSGP 1                & 220.57       &   *   & 01 04 33.30  -27 00 23.0 &     -   \\
CGG* SGP 51           & 384.71       &   *   & 01 04 28.60  -26 51 20.0 &    G5   \\
GSA 55                & 425.46       &   G   & 01 04 27.00  -27 03.50.0 &     -   \\ 
CGG* SGP 68           & 444.42       &   *   & 01 05 03.80  -26 51 45.0 &    G5   \\ 
RG 0102.6-2720        & 484.81       &   *   & 01 05.00.00  -27 04 00.0 &     -   \\               
\tableline
\end{tabular}
\end{center}
\end{table}

\begin{table}
\begin{center}
\caption{NED detection within a radius of 500 arcsec around the source position
of LBQS 0102-2713}
\begin{tabular}{llllll}
\tableline\tableline                                                                           
       Object Name                &  RA        &   DEC     &Type  &z   &dist\\
                                  &            &           &      &    & (arcmin) \\ 
    LBQS 0102-2713                &01h04m40.9s &-26d57m07s &QSO   &0.780000   &      0.0  \\
    2dFGRS S212Z160                &01h04m44.3s &-26d57m05s &G    &0.113844   &      0.8\\
    APMUKS(BJ)B010210.33-271359.0 &01h04m34.7s &-26d57m55s &G     &           &      1.6  \\
    2dFGRS S213Z276                &01h04m44.0s &-26d55m38s &G    &0.128000   &      1.6 \\
    APMUKS(BJ)B010215.35-271559.7 &01h04m39.7s &-26d59m55s &G     &           &      2.8 \\
    2dFGRS S213Z278                &01h04m28.9s &-26d55m40s &G    &0.127000   &      3.0 \\
    2MASX J01044626-2659591        &01h04m46.3s &-27d00m00s &G    &0.156813   &      3.1   \\
    APMUKS(BJ)B010217.75-271619.8 &01h04m42.1s &-27d00m16s &G     &           &      3.1 \\
    2MASX J01042726-2655252        &01h04m28.2s &-26d55m20s &G    &0.128500   &      3.4   \\
    2dFGRS S213Z275                &01h04m56.4s &-26d56m26s &G    &0.158043   &      3.5 \\
    APMUKS(BJ)B010212.20-271714.6 &01h04m36.6s &-27d01m10s &G     &           &      4.2  \\
    2dFGRS S212Z166                &01h04m20.1s &-26d56m42s &G    &0.129330   &      4.7\\
    APMUKS(BJ)B010210.57-271755.8 &01h04m34.9s &-27d01m51s &G     &           &      4.9\\
    APMUKS(BJ)B010241.39-271015.9 &01h05m05.7s &-26d54m12s &G     &           &      6.3 \\
    APMUKS(BJ)B010215.35-271953.3 &01h04m39.7s &-27d03m49s &G     &           &      6.7  \\
    2dFGRS S212Z165                &01h04m28.9s &-26d50m54s &G    &0.112910   &      6.8  \\
    2dFGRS S212Z164                &01h04m32.8s &-26d50m11s &G    &0.128985   &      7.2\\
    APMUKS(BJ)B010245.29-270904.2 &01h05m09.6s &-26d53m01s &G     &           &      7.6 \\
    APMUKS(BJ)B010251.36-271413.7 &01h05m15.7s &-26d58m10s &G     &           &      7.8  \\
    MDS ua-01-09                   &01h04m35.0s &-27d04m54s &G    &0.611600   &      7.9\\
    APMUKS(BJ)B010224.51-270526.6 &01h04m48.9s &-26d49m23s &G     &           &      8.0  \\
    APMUKS(BJ)B010252.42-271158.3 &01h05m16.7s &-26d55m55s &G     &           &      8.1\\
\tableline
\end{tabular}
\end{center}
\end{table}

\begin{figure}
\includegraphics[angle=270,scale=.300]{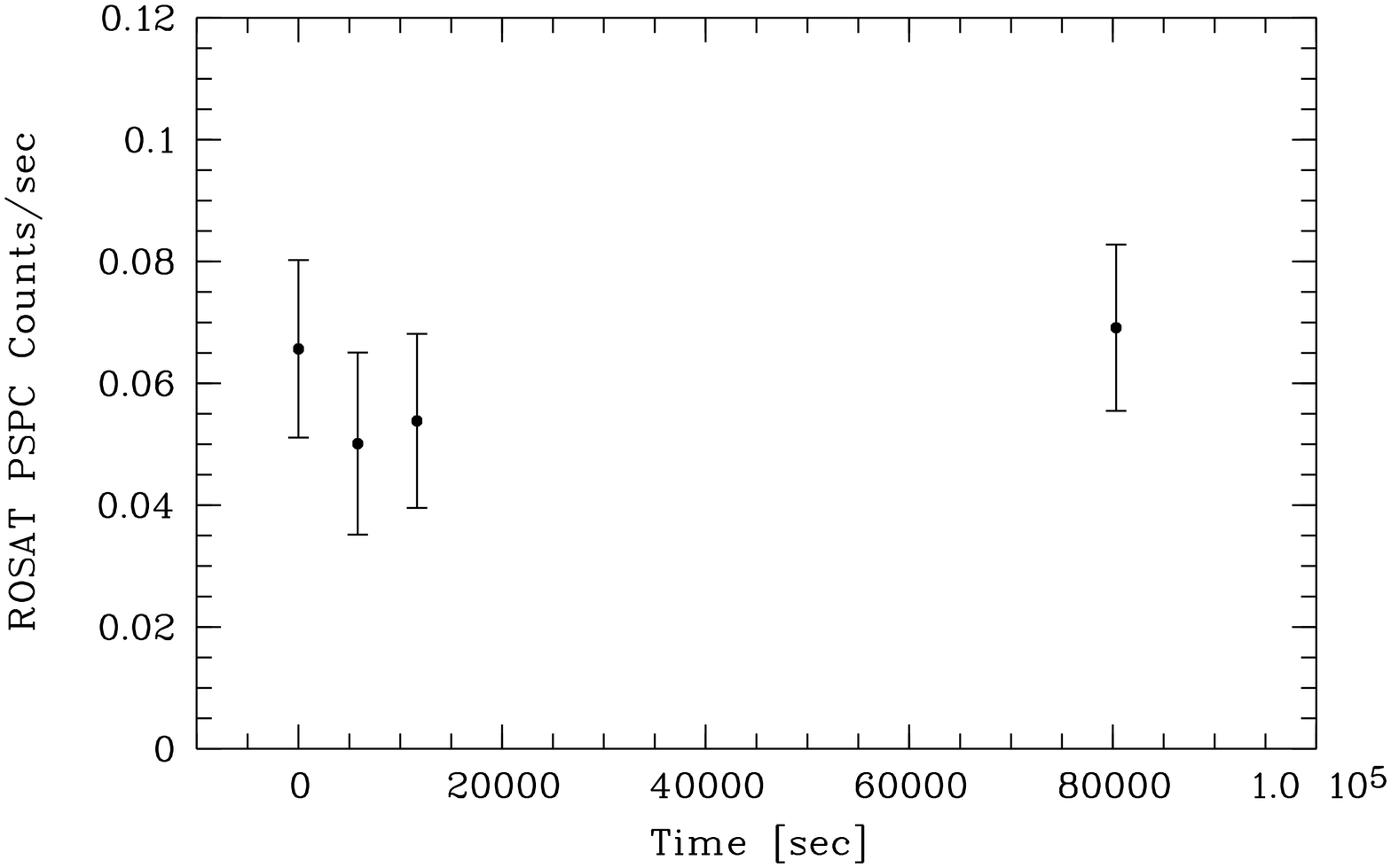}
\includegraphics[angle=270,scale=.300]{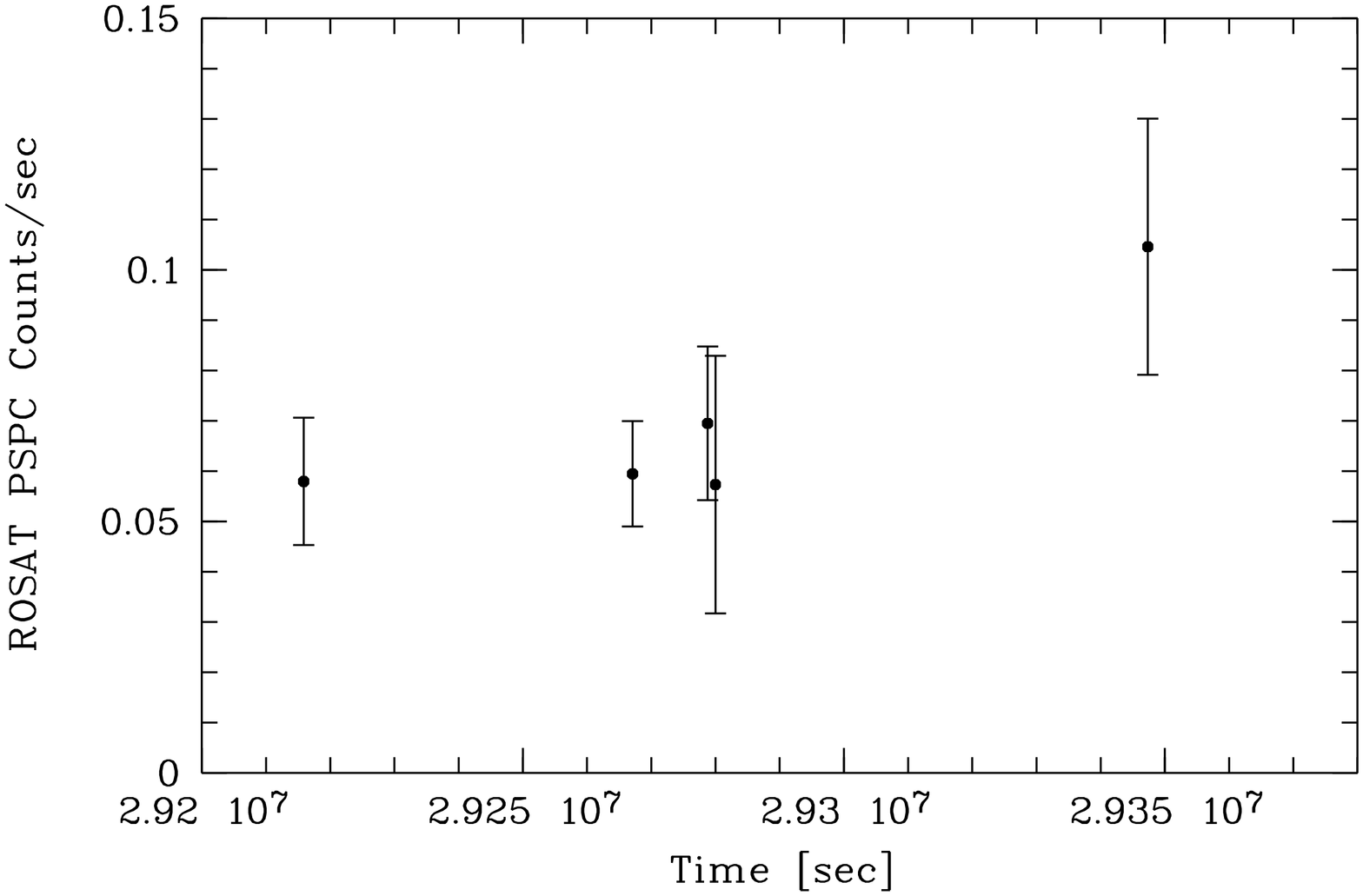}
\caption{ROSAT PSPC light curves of LBQS 0102-2713 observed
in January 1992 (left panel) and December 1992 (right panel).
The bin size is 4000 seconds. The source shows no significant
X-ray variability between the two observations.
 }
\end{figure}

\begin{figure}
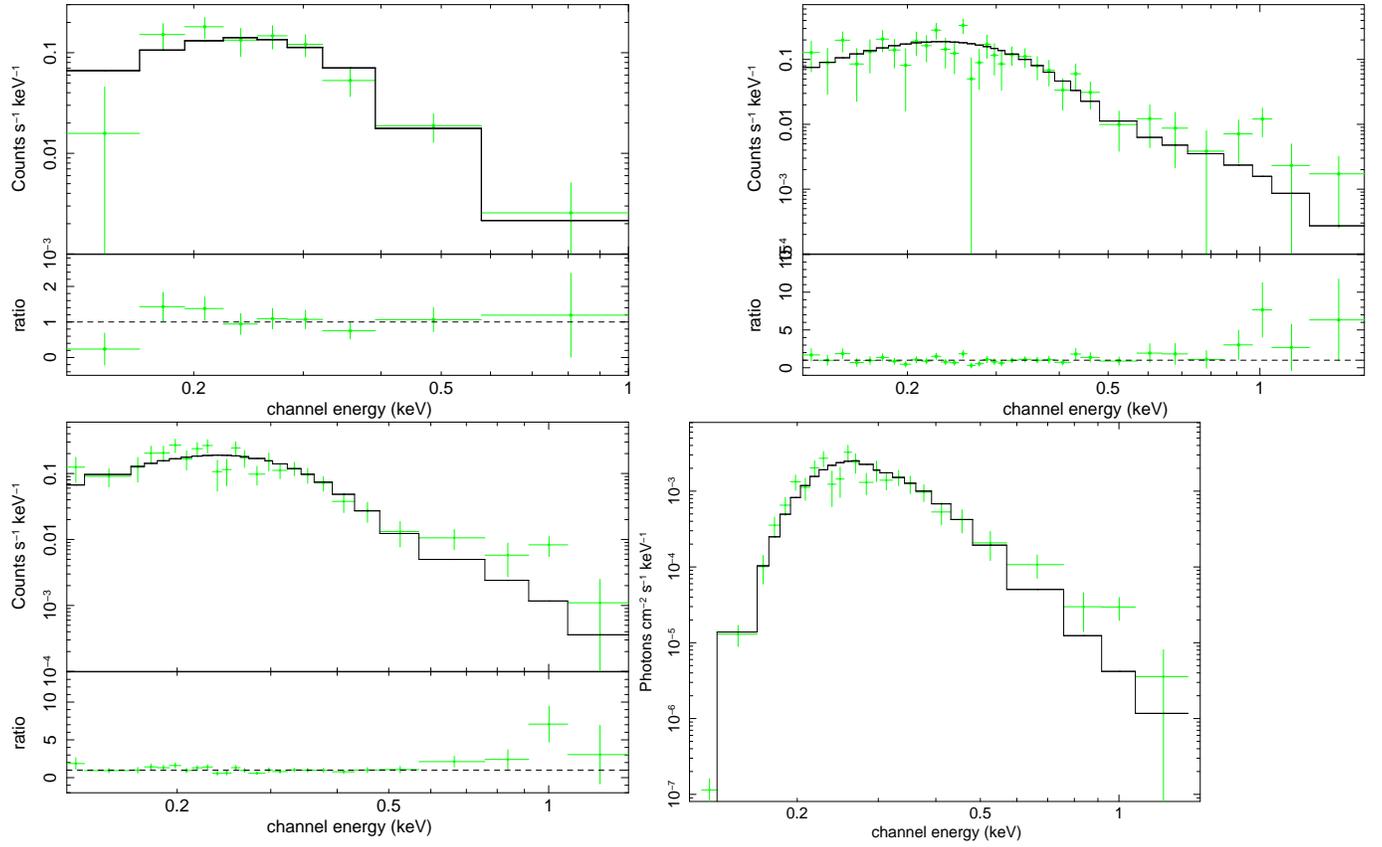

\includegraphics[angle=270,scale=.350]{fig2a.ps}
\includegraphics[angle=270,scale=.350]{fig2b.ps}
\includegraphics[angle=270,scale=.350]{fig2c.ps}
\includegraphics[angle=-90,width=7.5cm]{fig2d.ps}
\caption{ROSAT PSPC observation of LBQS 0102-2713 from January 1992 and
December 1992 (upper left and right pannels).
The lower left panel shows the fit to the merged data set.
 Applying a simple power-law fit and leaving the absorption 
 parameter free, we obtain from the {\it xselect} analysis
photon indices of 
 $\rm 6.6\pm2.5$, $\rm 5.5\pm2.5$ and $\rm 6.0\pm1.3$, respectively. 
 The unfolded spectrum obtained from the merged data set is shown in the
 lower right panel.
 }
\end{figure}

\begin{figure}
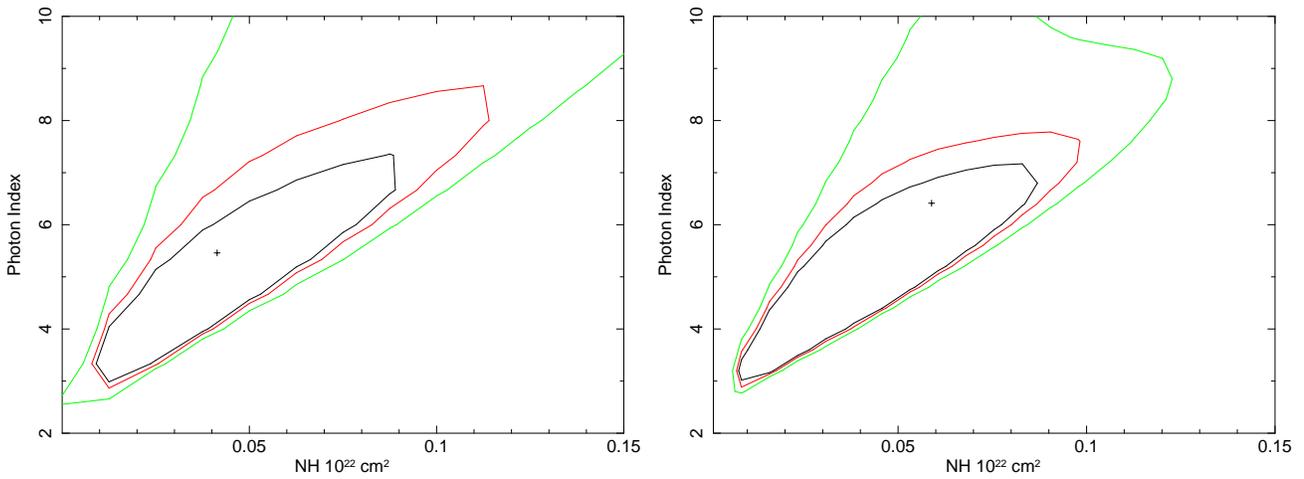

\includegraphics[angle=270,scale=.350]{fig3a.ps}
\includegraphics[angle=270,scale=.350]{fig3b.ps}
\caption{
$\rm \Gamma-N_H$ contour plots obtained from ROR number 700121p-0 (left) and
from the merged data set (right panel).
 The contour lines correspond to
68, 90, and 99 per cent confidence levels.
The low Galactic $\rm N_H$ value of $\rm 1.2\times 10^{20}\ cm^{-2}$
is within the 99 per cent confidence level resulting into an photon index of
about 3.5. The photon index might therefore be  flatter as indicated
by the power law fits and LBQS 0102-2713 might be less special in X-rays.
}
\end{figure}

\begin{figure}
\includegraphics[angle=0,scale=0.8,clip=]{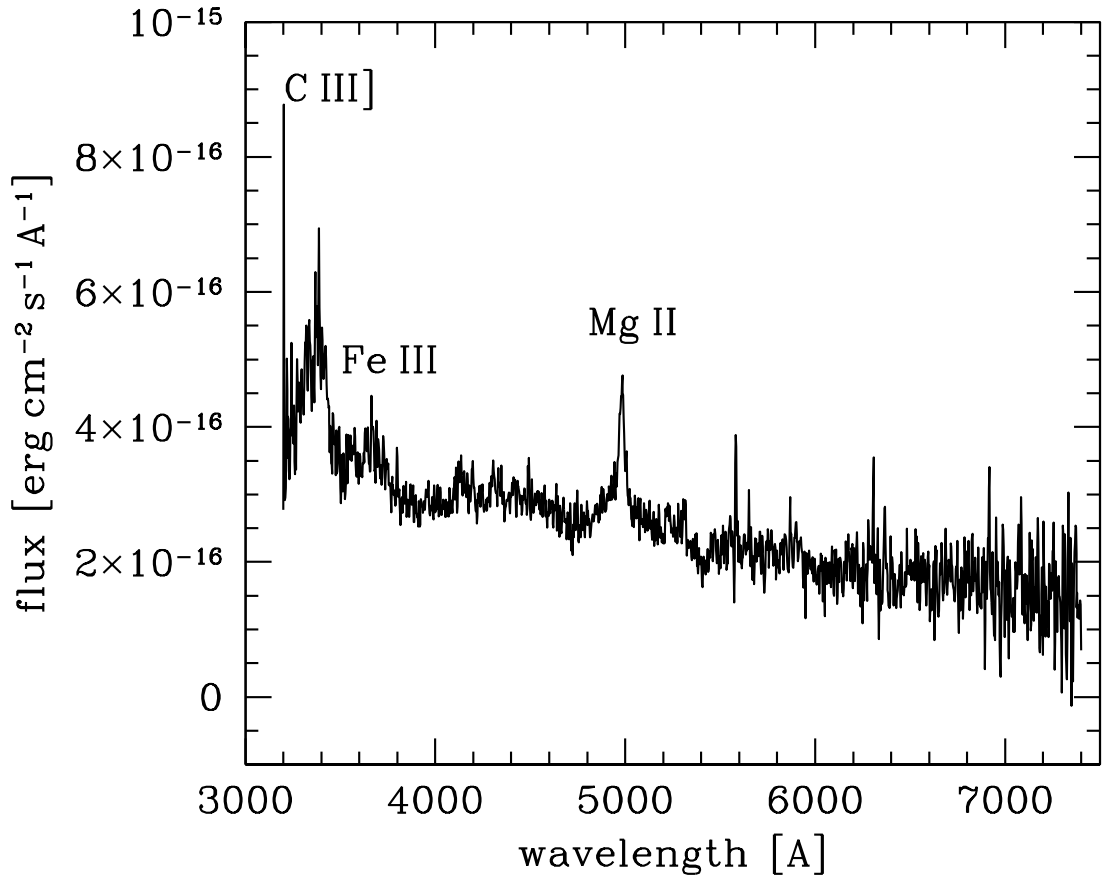}
\caption{
Optical spectrum of the quasar LBQS 0102-2713 obtained by Morris et al. (1991).
obtained in 1988. 
The strongest emission lines of C III], Fe III, and Mg II are marked. Between the Fe III
and the Mg II emission there is a strong Fe II UV multiplet emission between about 
2200 and 2500 \AA\ rest frame wavelength,  first noted by Morris et al.
(1991).}
\end{figure}

\begin{figure}
\includegraphics[angle=0,scale=.80,clip=]{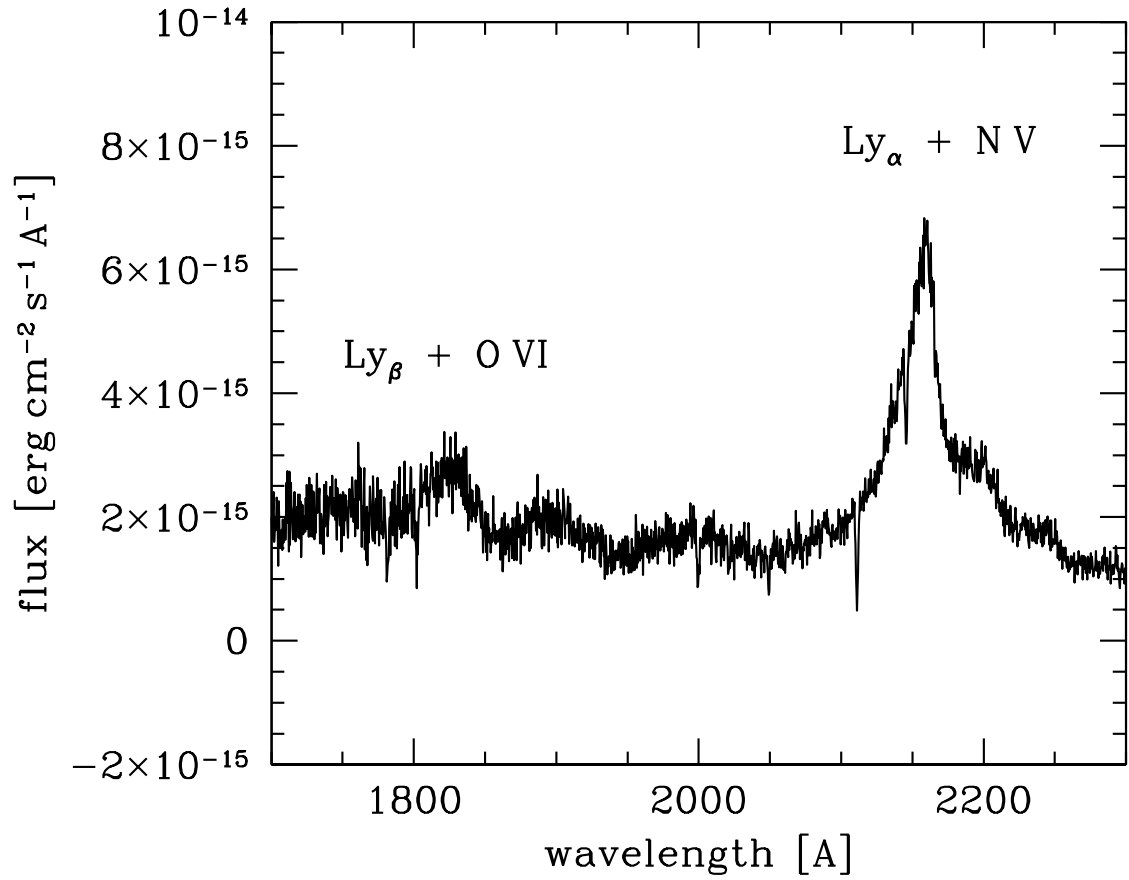}
\caption{
 HST FOS spectra of the quasar LBQS 0102-2713 obtained from  data set Y30901012T
 obtained in 1995. The strongest UV emission lines are marked and their EW values
 are comparable to those found in composite quasar spectra (c.f. Section 6.1).
 }
\end{figure}

\begin{figure}
\includegraphics[angle=-90,scale=.7,clip=]{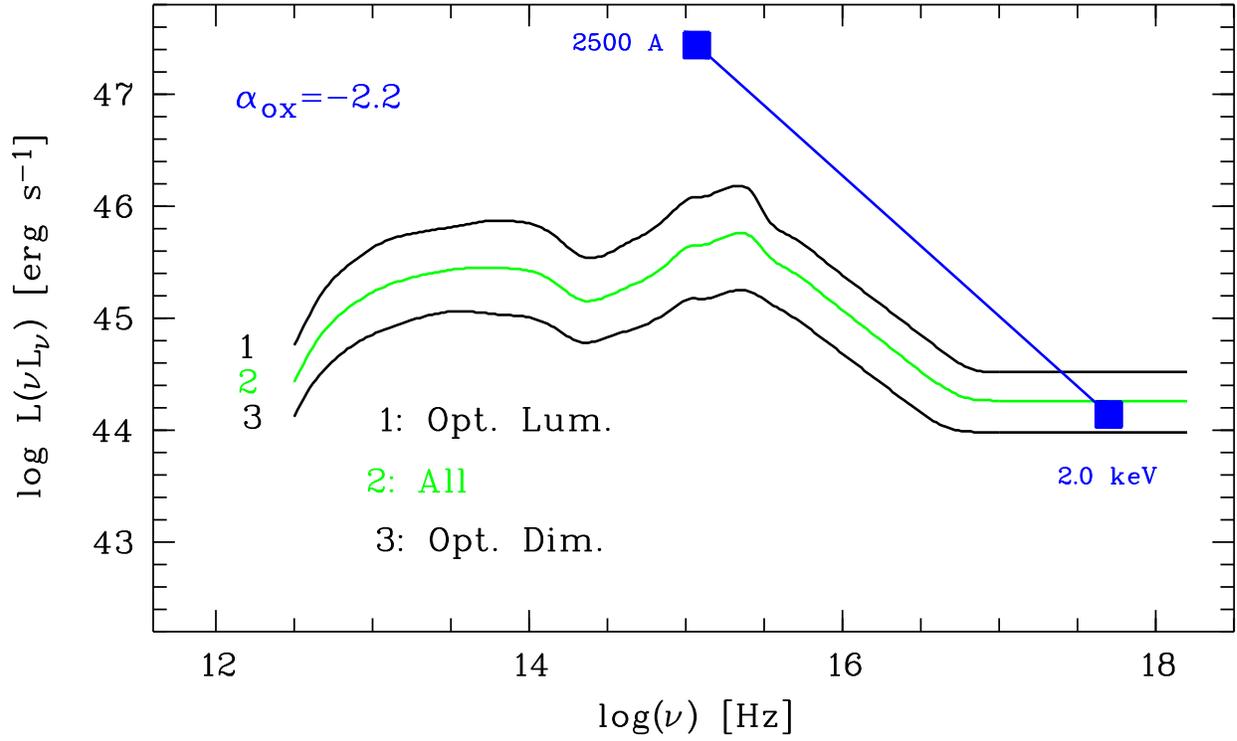}
\caption{
Comparision of the $\rm \nu L_{\nu}$ values in the UV and in the X-rays with 
the mean SED's for SDSS quasars. While the object is UV bright compared to the 
mean of the most luminous SDSS quasars (although there might be large scatter in
the individual luminosities), the 2 keV value is comparable 
to the SDSS dim quasars.
 }
\end{figure}

\begin{figure}
\includegraphics[angle=-90,scale=0.55,angle=90,clip=]{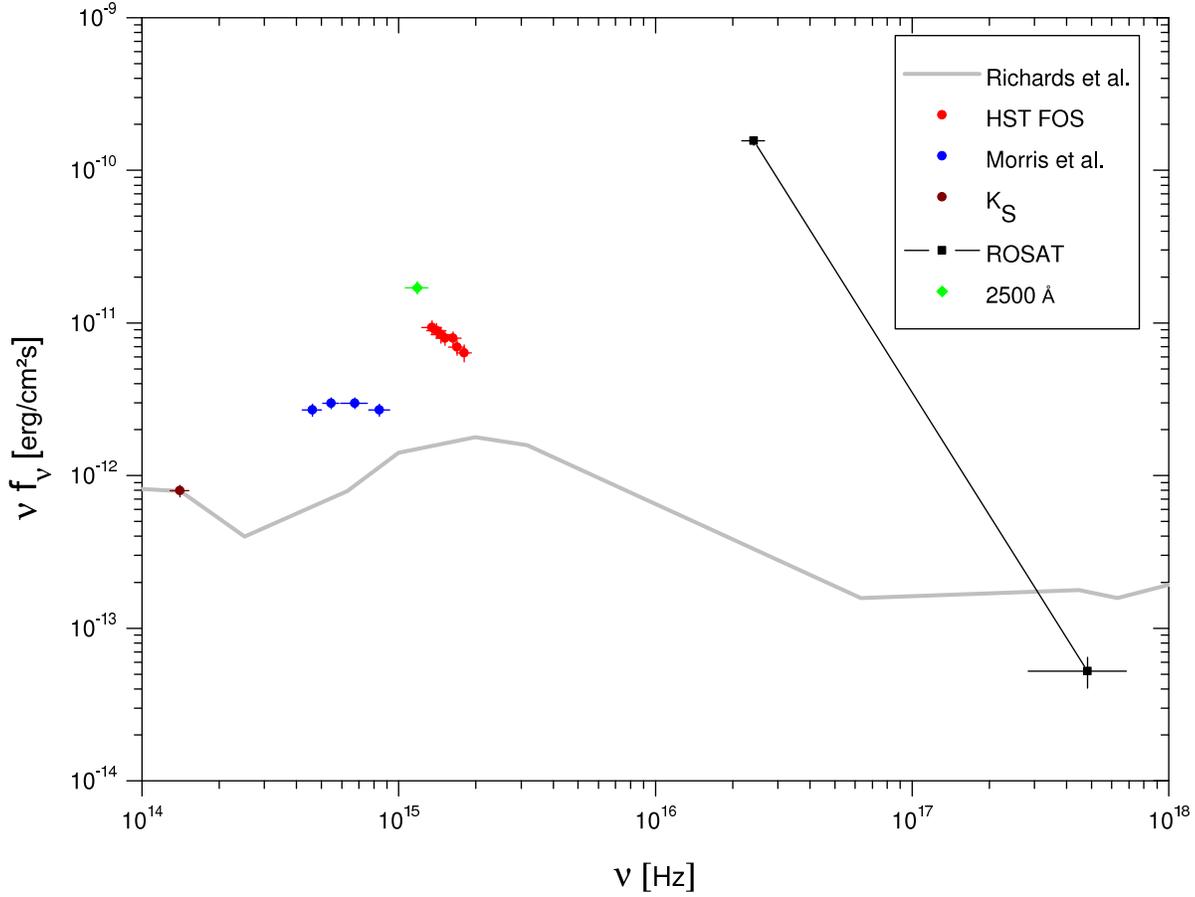}
\caption{
Optical spectrum from Morris et al. (1991), the HST FOS data,
the  ROSAT data and the $\rm K_S$ value in the rest frame.
The SED of LBQS 0102-2713 is normalized at the frequency of
the $\rm K_S$ $\rm \nu f_{\nu}$ value.
The unnormalized $\rm \nu f_{\nu}$ value
and the normalized one are $\rm 5.6\times10^{-13}\ erg\ cm^{-2}\ s^{-1}$
and $\rm 7.9\times10^{-13}\ erg\ cm^{-2}\ s^{-1}$, resulting into a
normalization factor of 1.4.
While in 
the UV the object is bright compared to the mean, at X-rays
the object is X-ray weak. 
We find a strong
discrepancy between the X-ray and the HST FOS UV data.
The data obtained from the optical spectrum are most probably
dominated by the host galaxy. 
 }
\end{figure}

\begin{figure}
\includegraphics[angle=0,scale=.90,clip=]{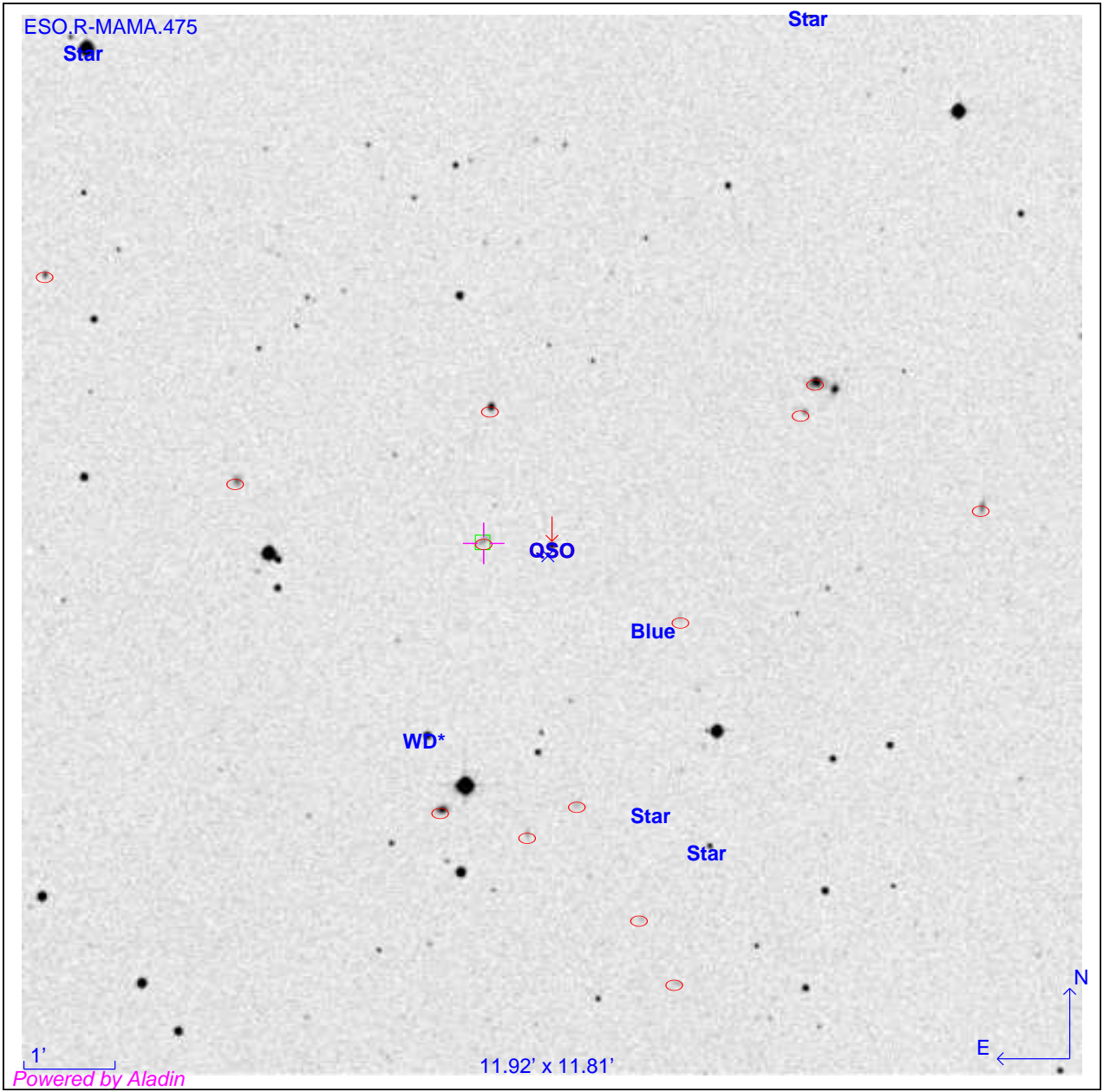}
\caption{ESO R-MAMA.475 image with SIMBAD and NED detections
overlaid. The SIMBAD objects are blue coloured and the red circles
indicate the NED detections. The box size is 10 times 10 arcmin.
LBQS 0102-2713 is located in the center.
 }
\end{figure}

\end{document}